\documentclass[conference]{IEEEtran}
\IEEEoverridecommandlockouts

\usepackage{graphicx}
\usepackage{booktabs}
\usepackage{amsmath,amssymb}
\usepackage{xcolor}
\usepackage{url}
\usepackage{cite}
\usepackage{balance}
\usepackage{multirow}
\usepackage{booktabs,threeparttable}
\usepackage[hidelinks]{hyperref}

\newcommand{\libname}{MultiFloats.cpp}
\newcommand{\mfjl}{MultiFloats.jl}
\newcommand{\ffx}[1]{\texttt{Float64x#1}}
\newcommand{\fsx}[1]{\texttt{Float32x#1}}
\newcommand{\nsop}{\,\text{ns/op}}
\newcommand{\mops}{\,\text{MOps/s}}

\begin{document}

\title{Portable Extended-Precision Floating-Point Arithmetic
for Scientific Computing Kernels}

\author{%
\IEEEauthorblockN{Manas Vishal}
\IEEEauthorblockA{%
\textit{Department of Mathematics}\\
\textit{University of Massachusetts Dartmouth}\\
North Dartmouth, MA, USA\\
vishalmanas28@gmail.com
}
}

\AddToHookNext{shipout/foreground}{%
\put(\dimexpr1in+\oddsidemargin\relax,\dimexpr-\paperheight+16pt\relax){%
\parbox[b]{\textwidth}{\scriptsize
\noindent M.~Vishal, ``Portable Extended-Precision Floating-Point Arithmetic
for Scientific Computing Kernels,'' in \textit{2026 IEEE High Performance
Extreme Computing Conference (HPEC)}, 2026.\par
\smallskip
\noindent \copyright~2026 IEEE. Personal use of this material is permitted.
Permission from IEEE must be obtained for all other uses, in any current or
future media, including reprinting/republishing this material for advertising
or promotional purposes, creating new collective works, for resale or
redistribution to servers or lists, or reuse of any copyrighted component of
this work in other works.
}%
}%
}

\maketitle

\begin{abstract}
Many scientific computations need more accuracy than IEEE double precision,
but arbitrary precision is often too costly on CPUs and GPUs. Multi-float
arithmetic fills this gap by representing each value as an unevaluated sum of
IEEE numbers. Existing implementations are generally confined to a CPU, CUDA,
or language-specific programming model.

We present \libname{}, a header-only C++17 library that uses Kokkos to compile
the same extended-precision kernels for Serial, OpenMP, CUDA, and HIP. The
library supports \ffx{2} through \ffx{4} (106--212 bits) and the corresponding
Float32-based types. We evaluate it using seven
SHOC-style kernels and compare against QD, CAMPARY, MPFR, and Julia \mfjl{} on
A100, GH200, and AMD RX~780M systems. On A100, \ffx{4} addition sustains
15,700~MOps/s, or $1.5\times$ CAMPARY; on CPU, vectorized \ffx{4} operations
are $4.2\text{--}6.6\times$ faster than QD's \texttt{qd\_real}. Addition and
multiplication produce 0~ULP error over
10,000 inputs, while \ffx{4} division and square root remain within 1.71 and
17.9~ULP, respectively. Thus, extended precision can remain practical and
portable without separate backend implementations.
\end{abstract}


\section{Introduction}
\label{sec:intro}

The accuracy of a scientific computation is often limited not by its
discretization, but by cancellation and the accumulation of floating-point
roundoff. The 53-bit significand of IEEE~754 \texttt{Float64}, for example,
cannot resolve intermediate quantities that cancel across more than roughly
21 orders of magnitude. Rump's polynomial~\cite{rump1988algorithms},
$f(a,b)=333.75b^6+a^2(11a^2b^2-b^6-121b^4-2)+5.5b^8+a/(2b)$, provides a
representative example. At $(77617,33096)$, \texttt{Float64} returns
$-1.18\times10^{21}$ although the true value is approximately $-0.8274$.
Even \ffx{2} (106 bits) gives the wrong sign, whereas \ffx{4} (212 bits)
recovers the correct result. Compensated summation can repair this particular
expression, but a general scientific code requires arithmetic that remains
accurate throughout the computation. These limitations are not confined to synthetic examples: our
gravitational-wave solvers for extreme-mass-ratio inspiral and self-force
computations have likewise required precision beyond
\texttt{Float64}~\cite{vishal2024toward,vishal2025superconvergent,vishal2026pnp}.

MPFR~\cite{fousse2007mpfr} provides arbitrary precision, commonly at
$10\text{--}100\times$ the cost of hardware arithmetic. Multi-float
arithmetic~\cite{bailey2005high} offers a complementary approach: a value is
stored as an unevaluated sum of $k$ IEEE numbers, yielding approximately
$53k$ bits for double-based types, while each operation is expressed in native
floating-point instructions. QD~\cite{hida2001qd},
CAMPARY~\cite{joldes2017campary}, and \mfjl{}~\cite{multifloatsjl} have shown
that this representation can be practical on CPUs, CUDA GPUs, and in Julia.
What remains missing is a compiled C++ implementation that uses one source
path across both NVIDIA and AMD GPUs.

We address this gap with \libname{}, a header-only C++17 library built on
Kokkos~\cite{trott2022kokkos}. Backend selection is made at configuration
time, while the arithmetic types and kernels remain unchanged. The main
contributions of this work are:

{\setlength{\itemsep}{0pt}\setlength{\parsep}{0pt}
\begin{enumerate}
  \item \textbf{Explicit SIMD types for extended precision on CPUs.}
        We introduce \texttt{Vec<T,W>}, a struct-of-arrays representation that
        exposes independent multi-float operations to the compiler. This
        avoids GCC's failure to vectorize the corresponding array-of-structures
        two-sum graph and improves add, multiply, and higher-limb division by
        $1.9\text{--}4.7\times$ over GCC scalar code.
  \item \textbf{A single source path for CUDA and HIP.}
        The same Kokkos kernels compile for NVIDIA and AMD GPUs. A direct
        comparison with raw CUDA measures Kokkos overhead at no more than
        1.7\% on A100 and GH200.
  \item \textbf{A verified extended-precision benchmark suite.}
        We evaluate seven SHOC-style kernels across \ffx{1}--\ffx{4} and
        \fsx{2}--\fsx{4}, with ULP measurements against MPFR and performance
        comparisons against QD, CAMPARY, and \mfjl{}.
\end{enumerate}}

\section{Background}
\label{sec:background}
\textbf{Multi-float arithmetic.}
We begin with $x=x_h+x_l$, where
$|x_l|\leq\frac{1}{2}\operatorname{ulp}(x_h)$; quad-double extends this
expansion to four limbs. \texttt{TwoSum} returns the rounded sum and its exact
residual, and \texttt{TwoProd} does the same for a
product~\cite{hida2001qd,dekker1971floating,joldes2017bounds}. Renormalization then
produces the observed 0-ULP addition and multiplication results. Cost grows
approximately as $O(k^2)$; on a GPU, larger $k$ increases both operations and
bytes per value, effects we separate in Section~\ref{sec:results-gpu}.

\textbf{Performance portability.}
Kokkos provides parallel loops, reductions, and views for Serial, OpenMP,
CUDA, HIP, and SYCL~\cite{trott2022kokkos}. Each kernel uses a
\texttt{KOKKOS\_LAMBDA}, so its C++ source is independent of the selected
execution space.

\textbf{Benchmark hierarchy.}
SHOC organizes accelerator workloads by algorithmic complexity
~\cite{danalis2010shoc}. AXPY, DOT, Reduction, Triad, and FFT expose arithmetic
and memory throughput, while GEMM and SpMV add reuse and irregular access. We
use this range to study extended precision beyond isolated operations.

\section{Library Implementation}
\label{sec:design}

\textbf{Core Types and Primitives.}
The fixed-size aggregate \texttt{MultiFloat<T,N>} stores $N$ IEEE limbs without
allocation or runtime polymorphism; \ffx{2}, for example, aliases
\texttt{MultiFloat<double,2>}. \texttt{KOKKOS\_INLINE\_FUNCTION} makes every
operator and mathematical function available on host and device.

The arithmetic follows Hida~\cite{hida2001qd}. Knuth's addition/subtraction
sequence makes \texttt{two\_sum} insensitive to FMA contraction, while
\texttt{two\_prod} computes its residual as \texttt{fma(a,b,-prod)} instead of
Dekker splitting. Thus, \libname{} can use \texttt{-ffp-contract=fast}, which
CAMPARY's splitting sequence cannot. Fixed two-sum sweeps renormalize each
expansion; five sweeps for \ffx{4} control GPU registers without affecting the
observed ULP error.

Newton--Raphson iterations implement division and square root. The default
\ffx{4} square root skips the three-limb stage in a $1{\to}2{\to}4$ refinement,
giving a maximum observed error of 17.9~ULP. \texttt{sqrt\_accurate} extends the reciprocal-
root refinement to four limbs and reaches a 1.2~ULP maximum at roughly twice the cost
(Section~\ref{sec:results-accuracy}).

\textbf{Kokkos Integration.}
Device-callable operations let multi-float values reside directly in a
\texttt{Kokkos::View}; extended-precision AXPY therefore retains the native
\texttt{parallel\_for} form. Changing only the CMake backend compiles this
source for Serial, OpenMP, CUDA, or HIP. Same-session raw-CUDA comparisons put
the abstraction cost below 1.7\% on A100 and GH200
(Section~\ref{sec:results-slowdown}).

\textbf{Vectorized CPU Types.}
An array-of-structures layout obscures independent limb operations from GCC.
Julia's LLVM pipeline appears to recover packed AVX instructions, whereas GCC
generally does not from the inlined two-sum graph. We expose this parallelism
through \texttt{Vec<T,W>}, which stores $W$ expansions in struct-of-arrays
order. On AVX2, \texttt{Vec4<Float64x2>} advances four values without a runtime
transpose, improving \ffx{2} addition and multiplication by
$1.9\text{--}2.2\times$ and remaining competitive with Julia at higher limb
counts (Section~\ref{sec:results-cpu}).

\begin{table}[t]
\centering
\caption{Benchmark parameters. Tables report medians of three independent
run-level best-of-five measurements.}
\label{tab:methodology}
\scriptsize
\setlength{\tabcolsep}{1pt}
\begin{tabular}{@{}l r r l l@{}}
\toprule
Benchmark  & $N$     & Reps$^*$ & Metric   & Layout \\
\midrule
CPU micro  & 10K     & 1    & ns/op    & AoS \\
GPU micro  & 10M     & 50   & MOps/s   & AoS \\
Level-1    & 1M      & 10   & MOps/s   & AoS \\
GEMM       & 64--4096 & 1    & GFLOPS   & LayoutLeft \\
SpMV       & 100K    & 10   & MOps/s   & CSR (5 nnz/row) \\
FFT        & 256--1M & 1    & MOps/s   & AoS (complex) \\
Accuracy   & 10K     & 1    & ULP stats & host \\
\bottomrule
\multicolumn{5}{@{}l}{\scriptsize $^*$Kernel invocations per timed trial.}
\end{tabular}
\end{table}

\section{Experimental Results}
We now evaluate the library from individual arithmetic operations through
multi-core and GPU kernels. Table~\ref{tab:methodology} summarizes the seven
SHOC-style workloads and the parameters used in each experiment.
\label{sec:results}

\subsection{Experimental Setup}
\label{sec:setup}

\textbf{Hardware and software.}
Our primary system contains an AMD EPYC~7313 16-core CPU (AVX2+FMA) and an
NVIDIA A100 80\,GB PCIe GPU. We compile the CPU code with GCC~13.3.0 and
Clang~21.1.8 using \texttt{-O3 -march=native -ffp-contract=fast}; the EPYC
frequency ceiling is 3.0\,GHz. The software stack includes CUDA~12.9
(NVHPC~25.7), Kokkos~4.7.02, Julia~1.12.5 with \mfjl{}~v3.2.6 and
CUDA.jl~6.2.0, QD~v2.3.22 with \texttt{QD\_IEEE\_ADD}, and MPFR as the
accuracy reference. We separately test HIP portability on an RX~780M
(RDNA~3) with ROCm~6.3.1.

\textbf{CPU protocol.}
For CPU micro-operations, we pin execution to one core with
\texttt{taskset~-c~0}, perform a three-second compute warmup, and time five
trials with $N=10{,}000$. The OpenMP experiments use $N=1$M, close thread
binding, 1--16 cores, and ten kernel invocations per trial. We report the
median of three independent runs.

\textbf{GPU protocol.}
For GPU micro-operations, we use $N=10$M, one warmup, 50 invocations per
trial, explicit fences, and five trials. The kernel experiments use $N=1$M
for Level~1, 100K rows for SpMV, and matrix sizes 64--4096 for GEMM. We verify
device quiescence with \texttt{nvidia-smi} and again report medians of three
independent runs.

\textbf{Inputs.}
The C++ and Julia performance tests use identical deterministic inputs, while
the accuracy tests use seeded random values. We archive each run-level range;
the median aggregation reduces sensitivity to CPU frequency ramping and the
launch-latency variation of small GPU workloads.

\subsection{CPU Micro-Operation Latency}
\label{sec:results-cpu}

We first isolate the cost of the arithmetic itself. Table~\ref{tab:cpu-micro}
compares scalar and Vec4 \libname{} operations with QD, \mfjl{}, and MPFR.

\begin{table}[t]
\centering
\caption{CPU micro-operation latency (ns/op). $N\!=\!10{,}000$, AMD EPYC~7313,
single core, turbo-warmed. GCC and Clang denote scalar code from GCC~13.3 and
Clang~21.1.8; Vec4 is the GCC-compiled SoA path. Bold: best (fastest) in each
column per precision class.}
\label{tab:cpu-micro}
\small
\setlength{\tabcolsep}{3.5pt}
\begin{tabular}{@{}ll rrrr@{}}
\toprule
Type & Mode & Add & Mul & Div & Sqrt \\
\midrule
\ffx{1}                  & GCC    &  \textbf{0.21} &  \textbf{0.21} &   \textbf{0.30} &    \textbf{2.29} \\
\ffx{1}                  & Clang  &  \textbf{0.21} &  0.22 &   0.31 &    2.42 \\
Julia \ffx{1}            & ---    &  \textbf{0.21} &  \textbf{0.21} &   0.31 &   2.30 \\
MPFR 53-bit             & ---    &  8.01 & 11.47 &  15.49 &   15.83 \\
\midrule
\ffx{2}                  & GCC    &  1.42 &  1.02 &   2.39 &    \textbf{3.67} \\
\ffx{2}                  & Clang  &  1.25 &  0.90 &   \textbf{1.33} &    3.70 \\
\ffx{2}                  & Vec4   &  \textbf{0.73} &  \textbf{0.47} &   3.58 &    3.71 \\
Julia \ffx{2}            & ---    &  1.25 &  0.93 &   5.18 &   19.28 \\
\texttt{dd\_real} (QD)  & ---    &  1.53 &  2.36 &   6.24 &    8.32 \\
MPFR 128-bit            & ---    &  8.11 & 17.06 &  46.34 &  118.4 \\
\midrule
\ffx{3}                  & GCC    & 11.73 &  7.42 &  99.53 &  125.02 \\
\ffx{3}                  & Clang  & 10.98 &  6.42 &  99.16 &  \textbf{107.69} \\
\ffx{3}                  & Vec4   &  \textbf{2.65} &  \textbf{1.70} &  \textbf{24.22} &  109.01 \\
Julia \ffx{3}            & ---    &  3.57 &  2.48 &  25.72 &  117.38 \\
MPFR 160-bit            & ---    &  9.87 & 23.60 &  67.89 &  120.0 \\
\midrule
\ffx{4}                  & GCC    & 23.12 & 24.14 & 197.15 &  199.69 \\
\ffx{4}                  & Clang  &  6.30 &  6.30 &  42.69 &  195.65 \\
\ffx{4}                  & Vec4   &  \textbf{6.09} &  \textbf{5.37} &  \textbf{42.20} &  190.63 \\
Julia \ffx{4}            & ---    &  6.32 &  6.38 &  43.29 & 189.62 \\
\texttt{qd\_real} (QD)  & ---    & 38.85 & 35.51 & 218.17 &  802.0 \\
MPFR 256-bit            & ---    & 13.52 & 31.96 &  79.08 &  \textbf{165.2} \\
\bottomrule
\end{tabular}
\end{table}

\begin{table}[t]
\centering
\caption{CPU micro-operation latency (ns/op) for the \texttt{Float32}-based
types, same protocol as Table~\ref{tab:cpu-micro}. Bold: best (fastest) in
each column per precision class.}
\label{tab:cpu-micro-f32}
\small
\setlength{\tabcolsep}{3.5pt}
\begin{tabular}{@{}ll rrrr@{}}
\toprule
Type & Mode & Add & Mul & Div & Sqrt \\
\midrule
\fsx{1}                   & GCC    & 0.11 & \textbf{0.10} & \textbf{0.12} & 1.35 \\
\fsx{1}                   & Clang  & 0.11 & \textbf{0.10} & \textbf{0.12} & 1.35 \\
Julia \fsx{1}             & ---    & \textbf{0.10} & \textbf{0.10} & \textbf{0.12} & \textbf{1.34} \\
\midrule
\fsx{2}                   & GCC    & 0.70 & 0.53 & 2.11 & \textbf{2.29} \\
\fsx{2}                   & Clang  & 2.18 & 0.94 & \textbf{1.85} & 2.34 \\
\fsx{2}                   & Vec4   & \textbf{0.41} & \textbf{0.27} & 3.46 & 2.35 \\
Julia \fsx{2}             & ---    & 0.55 & 0.40 & 2.46 & 18.00 \\
\midrule
\fsx{3}                   & GCC    & 12.06 & 7.97 & 128.30 & \textbf{106.86} \\
\fsx{3}                   & Clang  & 10.56 & 6.75 & 104.22 & 112.23 \\
\fsx{3}                   & Vec4   & 2.86 & \textbf{1.77} & 23.96 & 107.62 \\
Julia \fsx{3}             & ---    & \textbf{2.48} & 1.95 & \textbf{13.81} & 112.21 \\
\midrule
\fsx{4}                   & GCC    & 21.97 & 22.61 & 191.70 & 191.74 \\
\fsx{4}                   & Clang  & 22.94 & 21.90 & 209.23 & 213.00 \\
\fsx{4}                   & Vec4   & 6.22 & 5.75 & 41.91 & 186.84 \\
Julia \fsx{4}             & ---    & \textbf{4.87} & \textbf{5.03} & \textbf{23.38} & \textbf{181.03} \\
\bottomrule
\end{tabular}
\end{table}

The measurements reveal that data layout and compiler code generation are as
important as the arithmetic algorithm. Vec4 improves addition,
multiplication, and three- or four-limb division by $1.9\text{--}4.7\times$
relative to GCC scalar code. The two-limb division is the exception: a
correctness-preserving GCC anti-vectorization guard raises Vec4 latency to
3.58\nsop{}, compared with 2.39\nsop{} for scalar GCC, although addition and
multiplication still benefit from the explicit layout.

At \ffx{4}, Vec4 requires 6.09, 5.37, and 42.20\nsop{} for addition,
multiplication, and division, closely matching or improving on Julia's
6.32, 6.38, and 43.29\nsop{}. GCC scalar requires 23.12, 24.14, and
197.15\nsop{}, whereas Clang recovers most of this gap. We therefore attribute
much of the scalar difference to code generation rather than to the
underlying representation. Square root benefits less because its Newton
iteration is sequential. Relative to QD, the Vec4 implementation is
$4.2\text{--}6.6\times$ faster across the four \ffx{4} operations.

The Float32-based results in Table~\ref{tab:cpu-micro-f32} follow the same
general pattern. At two limbs, Vec4 addition takes 0.41\nsop{}, compared with
0.73\nsop{} for the double-based type. At larger limb counts, however,
renormalization dominates and the Float32 and Float64 scalar costs approach
one another. Julia leads several three- and four-limb operations, while the
explicit Vec4 path remains strongest for three-limb multiplication.

\subsection{CPU Transcendentals}
\label{sec:results-trans}

Transcendental functions introduce argument reduction and polynomial
evaluation in addition to the multi-limb primitives. Table~\ref{tab:trans}
reports their CPU latency and A100 throughput.

\begin{table}[t]
\centering
\caption{Transcendental performance: CPU latency (ns/op, scalar) and
GPU throughput (MOps/s, $N\!=\!10$M, A100).}
\label{tab:trans}
\small
\setlength{\tabcolsep}{2.5pt}
\begin{tabular}{@{}l l rrrr@{}}
\toprule
Platform & Type & exp & log & sinpi & cbrt \\
\midrule
\multirow{6}{*}{CPU (ns/op)}
  & \ffx{2} & 186.6 & 141.9 & 227.6 & 79.04 \\
  & \ffx{3} & 873.9 & 625.4 & 822.7 & 259.7 \\
  & \ffx{4} & 2202  & 1425  & 1786  & 477.1 \\
  & \fsx{2} & 100.6 & 102.1 &  88.9 & 75.29 \\
  & \fsx{3} & 456.2 & 392.9 & 461.4 & 343.9 \\
  & \fsx{4} & 923.9 & 827.3 & 1074  & 468.8 \\
\midrule
\multirow{6}{*}{GPU (MOps/s)}
  & \ffx{2} & 10,952 & 2,025 & 10,945 & 24,937 \\
  & \ffx{3} &  2,181 &   979 &  2,262 &  6,706 \\
  & \ffx{4} &    757 &   557 &    768 &  2,992 \\
  & \fsx{2} & 35,223 & 29,494 & 38,591 & 46,952 \\
  & \fsx{3} &  7,640 &  2,084 &  8,272 & 13,399 \\
  & \fsx{4} &  2,802 &  1,235 &  2,804 &  5,985 \\
\bottomrule
\end{tabular}
\end{table}

We observe an effective $O(k^2)$--$O(k^3)$ cost as argument reduction and
Horner evaluation are repeated in extended precision. On the GPU,
\ffx{2} \texttt{exp} reaches 10,952~\mops{}, while \ffx{4} reaches
757~\mops{}. The resulting $14.5\times$ slowdown is larger than the
$4.3\times$ slowdown for addition, illustrating the growing cost of the
polynomial evaluation. Against 1024-bit MPFR, \texttt{exp}, \texttt{log},
and \texttt{sinpi} have maximum errors of 8--38~ULP and median errors below
one ULP across \ffx{2}--\ffx{4}. The shared functions agree with Julia;
our independently implemented \texttt{sinpi} shows the same error range.

\subsection{OpenMP Multi-Core Scaling}
\label{sec:results-omp}

We next examine whether the additional arithmetic per value remains scalable
across CPU cores. Using Kokkos::OpenMP with close thread binding, we run the
SHOC kernels on 1--16 EPYC~7313 cores. Table~\ref{tab:omp} reports \fsx{4}
throughput, while Table~\ref{tab:omp-eff} gives the 16-thread efficiency for
all eight precision levels.

\begin{table}[t]
\centering
\caption{OpenMP scaling: \fsx{4} throughput (MOps/s) on EPYC~7313.}
\label{tab:omp}
\small
\setlength{\tabcolsep}{5pt}
\begin{threeparttable}
\begin{tabular}{@{}l rrrr r@{}}
\toprule
Kernel & 1\,thr & 4\,thr & 16\,thr & Speedup & Eff. \\
\midrule
AXPY      & 17.2 &  68.1 & 259 & 15.1$\times$ & 94\% \\
DOT       & 17.3 &  67.6 & 256 & 14.8$\times$ & 93\% \\
Reduction & 27.1 & 106.5 & 402 & 14.8$\times$ & 93\% \\
Triad     & 17.1 &  68.0 & 258 & 15.1$\times$ & 94\% \\
SpMV      & 19.2 &  76.6 & 290 & 15.1$\times$ & 94\% \\
FFT       &  2.6 &  10.3 & 34.3 & 13.2$\times$ & 82\% \\
\bottomrule
\end{tabular}
\begin{tablenotes}[flushleft]
\footnotesize
\item $N\!=\!1$M; 10 invocations/trial, best of 5; median of 3 runs.
\end{tablenotes}
\end{threeparttable}
\end{table}

\begin{table}[t]
\centering
\caption{OpenMP 16-thread parallel efficiency ($S_{16}/16$, percent)
across kernels and precisions on the EPYC~7313.}
\label{tab:omp-eff}
\small
\setlength{\tabcolsep}{4pt}
\begin{threeparttable}
\begin{tabular}{@{}l rrrr rrrr@{}}
\toprule
 & \multicolumn{4}{c}{\texttt{Float64}$\times$} & \multicolumn{4}{c}{\texttt{Float32}$\times$} \\
\cmidrule(lr){2-5} \cmidrule(l){6-9}
Kernel & 1 & 2 & 3 & 4 & 1 & 2 & 3 & 4 \\
\midrule
AXPY      & 54 & 88 & 97 & 96 & 83 & 94 & 94 & 94 \\
DOT       & 85 & 93 & 93 & 93 & 86 & 92 & 91 & 93 \\
Reduction & 86 & 92 & 93 & 93 & 87 & 92 & 93 & 93 \\
Triad     & 46 & 93 & 95 & 95 & 57 & 93 & 94 & 94 \\
SpMV      & 78 & 92 & 93 & 93 & 79 & 92 & 94 & 94 \\
FFT       & 20 & 32 & 57 & 74 & 44 & 61 & 69 & 82 \\
\bottomrule
\end{tabular}
\begin{tablenotes}[flushleft]
\footnotesize
\item $N\!=\!1$M, 16 threads, close binding; median of 3 run-level
best-of-5 values.
\end{tablenotes}
\end{threeparttable}
\end{table}

At two or more limbs, AXPY, DOT, Reduction, Triad, and SpMV sustain
$88\text{--}97\%$ parallel efficiency. Multi-limb renormalization increases
the computation performed per byte, so kernels that are bandwidth-bound at
native precision move toward a compute-bound regime and scale more nearly with
core count. FFT exhibits the same trend: efficiency rises from 20 to 74\% for
the Float64 family and from 44 to 82\% for the Float32 family as the limb count
increases. We interpret this monotonic improvement as evidence that memory
traffic, rather than Kokkos synchronization overhead, limits the lower-
precision cases.

\subsection{GPU Throughput}
\label{sec:results-gpu}

We then repeat the micro-operation study on A100. Table~\ref{tab:gpu-micro}
compares the four basic operations with \mfjl{} and, for double-based types,
CAMPARY.

\begin{table}[t]
\centering
\caption{GPU throughput (MOps/s). $N\!=\!10$M elements on NVIDIA A100.
Bold: best in each precision class. CAMPARY is built on \texttt{double}
limbs, so the \texttt{Float32}-based rows compare \libname{} and Julia only.}
\label{tab:gpu-micro}
\small
\setlength{\tabcolsep}{3pt}
\begin{tabular}{@{}ll rrrr@{}}
\toprule
Type & Library & Add & Mul & Div & Sqrt \\
\midrule
\ffx{1} & MF.cpp   & \textbf{67,579} & \textbf{67,556} & \textbf{67,198} & \textbf{96,356} \\
\ffx{1} & Julia    & 64,568 & 65,045 & 64,978 & 93,522 \\
\midrule
\ffx{2} & MF.cpp   & \textbf{34,786} & \textbf{34,773} & \textbf{34,456} & \textbf{51,182} \\
\ffx{2} & Julia    & 34,338 & 34,414 & 34,202 & 47,601 \\
\ffx{2} & CAMPARY  & 34,238 & 34,231 & 34,127 & 49,282 \\
\midrule
\ffx{3} & MF.cpp   & \textbf{23,041} & \textbf{23,172} & \textbf{14,863} & \textbf{14,779} \\
\ffx{3} & Julia    & 22,927 & 23,021 & 14,317 & 13,899 \\
\ffx{3} & CAMPARY  & 16,466 & 16,713 & 10,001 &  8,969 \\
\midrule
\ffx{4} & MF.cpp   & 15,726 & 16,605 &  \textbf{7,099} &  \textbf{7,275} \\
\ffx{4} & Julia    & \textbf{16,581} & \textbf{17,189} &  6,977 &  7,067 \\
\ffx{4} & CAMPARY  & 10,468 & 10,520 &  6,601 &  6,468 \\
\midrule
\fsx{1} & MF.cpp & \textbf{120,837} & \textbf{120,843} & \textbf{119,254} & \textbf{160,214} \\
\fsx{1} & Julia  & 117,359 & 118,562 & 117,167 & 151,685 \\
\midrule
\fsx{2} & MF.cpp & \textbf{66,684} & \textbf{67,480} & \textbf{62,189} & \textbf{95,161} \\
\fsx{2} & Julia  & 64,382 & 64,648 & 61,249 & 72,572 \\
\midrule
\fsx{3} & MF.cpp & \textbf{44,939} & \textbf{45,513} & \textbf{28,052} & \textbf{28,517} \\
\fsx{3} & Julia  & 44,134 & 44,762 & 25,104 & 24,728 \\
\midrule
\fsx{4} & MF.cpp & 33,623 & 33,673 & \textbf{13,915} & \textbf{14,487} \\
\fsx{4} & Julia  & \textbf{33,816} & \textbf{33,826} & 12,994 & 13,199 \\
\bottomrule
\end{tabular}
\end{table}

At one and two limbs, addition and multiplication have an arithmetic
intensity of approximately 0.42~FLOP/byte, well below the A100 roofline knee
of 4.85~FLOP/byte. These operations are therefore bandwidth-bound, and the
three implementations agree within 2\% at \ffx{2}. Square root can exceed
addition in this regime because it reads one input array instead of two; the
difference reflects memory traffic rather than a cheaper algorithm.

As the limb count increases, arithmetic intensity exposes differences in the
implementations. At \ffx{4}, \libname{} is $1.1\text{--}1.6\times$ faster
than CAMPARY; addition, for example, reaches 15,726 rather than
10,468~\mops{}. We attribute much of this gap to the use of explicit FMA
residuals, which lets \libname{} retain fused operations while CAMPARY's
Dekker splitting requires \texttt{--fmad=false}. \libname{} and Julia remain
within 6\% for all four \ffx{4} operations, consistent with small differences
in instruction scheduling. A direct raw-CUDA comparison places the Kokkos
abstraction overhead below 1.7\% on both A100 and GH200.

\subsection{SHOC Kernel Performance}
\label{sec:results-kernels}

\begin{figure}[t]
  \centering
  \includegraphics[width=\columnwidth]{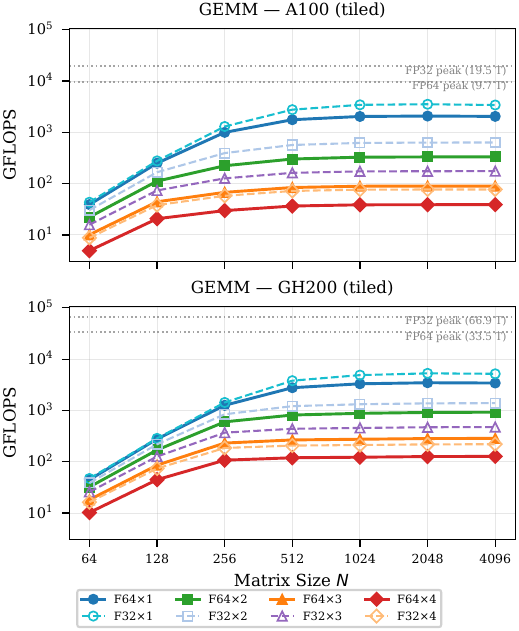}
  \caption{Tiled GEMM throughput scaling on A100 (top)
  and GH200 (bottom) for
  \texttt{Float64}- and \texttt{Float32}-based types across matrix sizes
  ($N\!=\!64$--$4096$).  Solid lines: Float64; dashed: Float32.
  Dotted grey lines show theoretical non-Tensor-Core FP64/FP32 peaks.
  Both GPUs use identical Kokkos source.}
  \label{fig:gemm}
\end{figure}

Micro-operation throughput alone does not describe how the arithmetic behaves
inside complete kernels. Figure~\ref{fig:gemm} and
Table~\ref{tab:kernels} therefore report all seven SHOC-style workloads across
the supported precision levels.

\begin{table}[t]
\centering
\caption{A100 GPU kernel throughput (MOps/s): $N\!=\!1$M for Level~1/FFT
and 100K rows for SpMV. At $N\!=\!1$M, low-limb working sets ($\leq24$\,MB;
48\,MB for \ffx{2}) fit within or border the 40\,MB L2 cache. This explains
the 2.1--2.5\,TB/s implied AXPY/Triad traffic above the 2.0\,TB/s HBM peak;
Table~\ref{tab:gpu-micro} gives HBM-resident $N\!=\!10$M rates.}
\label{tab:kernels}
\small
\setlength{\tabcolsep}{4pt}
\begin{tabular}{@{}l rrrr@{}}
\toprule
          & \multicolumn{4}{c}{\texttt{Float64}$\times$N} \\
\cmidrule(lr){2-5}
Kernel    & x1 & x2 & x3 & x4 \\
\midrule
AXPY      & 102,469 &  42,802 &  19,969 &  15,334 \\
DOT       &  41,384 &  27,526 &  14,491 &   8,334 \\
Reduction &  43,070 &  31,606 &  18,998 &  10,969 \\
Triad     & 102,637 &  29,311 &  19,832 &  15,102 \\
SpMV      &  73,001 &  50,190 &  27,383 &  15,508 \\
FFT       &   5,684 &   2,545 &   1,736 &   1,330 \\
\midrule
          & \multicolumn{4}{c}{\texttt{Float32}$\times$N} \\
\cmidrule(lr){2-5}
          & x1 & x2 & x3 & x4 \\
\midrule
AXPY      & 104,722 & 101,998 &  52,734 &  28,689 \\
DOT       &  43,786 &  35,565 &  22,601 &  14,181 \\
Reduction &  45,537 &  37,649 &  25,645 &  18,145 \\
Triad     & 104,569 &  99,630 &  36,276 &  25,792 \\
SpMV      &  77,952 &  64,573 &  42,596 &  25,716 \\
FFT       &   8,532 &   4,496 &   3,194 &   2,306 \\
\bottomrule
\end{tabular}
\end{table}

The kernel results separate memory-dominated and compute-dominated behavior.
AXPY and Triad incur an approximately $7\times$ \ffx{4}/\ffx{1} slowdown,
with part of the additional arithmetic hidden by memory access. DOT and
Reduction slow by $4\text{--}5\times$, and SpMV by $4.7\times$.

GEMM is the most sensitive to precision because every multiply--accumulate
executes the full multi-limb graph. At $N=4096$, the naive A100 kernel falls
from 283~GFLOPS at \ffx{1} to 38.8~GFLOPS at \ffx{4}. A $16\times16$
shared-memory tile raises \ffx{1} to 2,049~GFLOPS and \ffx{2} to
333~GFLOPS, but provides little benefit at three and four limbs, where the
$O(k^2)$ arithmetic dominates data reuse. On GH200, the corresponding tiled
rates are 3,423 and 127~GFLOPS for \ffx{1} and \ffx{4}. Against 1,024-bit
MPFR for $N\leq32$, GEMM remains below 10~ULP at \ffx{2} and produces
0.0~ULP error at \ffx{3}--\ffx{4}.

\textbf{FFT.}
For FFT, we precompute the extended-precision twiddle table rather than
evaluating \texttt{sincospi} in each butterfly. At $N=1$M, throughput decreases
from 5,684 to 1,330~\mops{} between \ffx{1} and \ffx{4}, a $4.3\times$
slowdown. Archived single-core measurements give GPU speedups that grow from
$74\times$ to $649\times$, since the GPU can hide more of the arithmetic as each butterfly
becomes more expensive. Comparison with an $O(N^2)$ MPFR DFT at $N=256$
gives mean errors of 5.4, 3.2, 1.5, and 1.0~ULP and maxima of 810, 365, 157,
and 100~ULP; both measures decrease with limb count.

The Float32-based representation offers a useful intermediate regime. On
bandwidth-bound AXPY, \fsx{2} reaches 101,998~\mops{}, or 99\% of native
\ffx{1}, while providing an approximately 48-bit significand. Its exponent
range remains that of Float32, so it is useful when precision and bandwidth
dominate but is not a general replacement for \texttt{Float64}.

\subsection{Slowdown and Overhead}
\label{sec:results-slowdown}

\begin{figure}[t]
  \centering
  \includegraphics[width=\columnwidth]{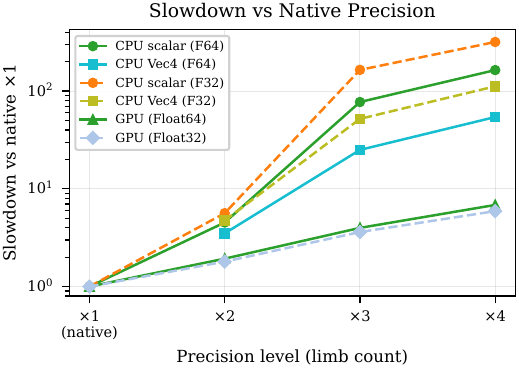}
  \caption{Slowdown versus native precision (\ffx{1} or \fsx{1})
  across CPU scalar, CPU Vec4 (both Float64-based), GPU Float64-based,
  and GPU Float32-based types.  Geometric mean over add, mul, div, sqrt.}
  \label{fig:slowdown}
\end{figure}

Figure~\ref{fig:slowdown} summarizes the cost relative to native precision.
On A100, double-based addition slows by $1.9\times$, $2.9\times$, and
$4.3\times$ at two, three, and four limbs. These ratios should not be read as
pure operation counts because the native baseline is bandwidth-limited; the
absolute rates in Table~\ref{tab:gpu-micro} provide the more meaningful GPU
comparison. Float32-based types have a four-limb geometric-mean slowdown of
$5.9\times$, consistent with the cost of single-precision FMA. On CPU, GCC
scalar slows by $87\text{--}657\times$ at \ffx{4}, while Vec4 and Clang reduce
this range to $26\text{--}141\times$. This difference again shows that the
observed cost includes a substantial compiler-code-generation component.

\subsection{Accuracy}
\label{sec:results-accuracy}

\begin{figure}[t]
  \centering
  \includegraphics[width=\columnwidth]{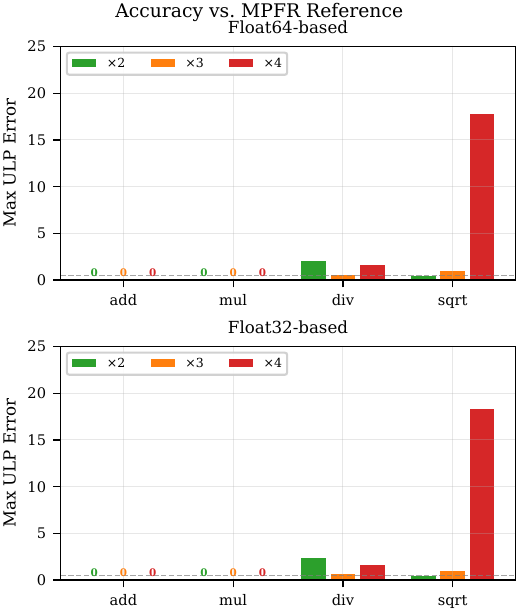}
  \caption{ULP error at target precision for \texttt{Float64}-based (top)
  and \texttt{Float32}-based (bottom) types versus 1024-bit MPFR reference.}
  \label{fig:accuracy}
\end{figure}

We evaluate \ffx{2}--\ffx{4} at their $53k$-bit target precision against a
1,024-bit MPFR reference using 10,000 random inputs in
$[0.1,1000]$ (Fig.~\ref{fig:accuracy}). Addition and multiplication give
0.0~ULP error at every tested precision, confirming the error-free
transformations. At \ffx{2}, division has a maximum error of 2.15~ULP
(median 0.11), while square root reaches 0.45~ULP (median 0.06). At
\ffx{4}, the corresponding maxima are 1.71 and 17.9~ULP, with medians of
0.02 and 0.14~ULP.

The larger \ffx{4} square-root maximum follows from the $1{\to}2{\to}4$
refinement: skipping the three-limb step approximately doubles throughput while
keeping the observed error below 18~ULP. \texttt{sqrt\_accurate} reduces this
maximum to 1.2~ULP at twice the cost. Without FMA contraction, \libname{} and
\mfjl{} produce bit-identical \ffx{3}--\ffx{4} outputs. The Float32-based
division/square-root maxima are 2.40/0.46~ULP at two limbs and
1.72/19.85~ULP at four.

Two limitations follow from the underlying IEEE representation. Subnormal
inputs can underflow intermediate products, and values near the Float64
overflow threshold degrade the square-root refinement. We also find a
compiler-specific sensitivity: Clang~21 SLP vectorization corrupts the low
limbs of the \ffx{2} FFT under the reported optimization flags. We therefore
disable SLP only for Clang FFT targets. The corrected Clang build passes the
MPFR accuracy test, and all kernel throughput reported here uses GCC on CPU or CUDA on
GPU.

\subsection{Cross-Vendor Portability}
\label{sec:results-portability}

Finally, we test portability beyond compilation. The complete suite runs on an
AMD RX~780M through Kokkos::HIP and ROCm~6.3.1 by changing only the CMake
backend. Table~\ref{tab:portability} compares representative \ffx{2} CUDA and
HIP results. Although the devices belong to different hardware classes, the
same arithmetic and kernels execute in both environments.

\begin{table}[t]
\centering
\caption{Cross-vendor portability: \ffx{2} throughput (MOps/s)
from identical source on NVIDIA CUDA vs.\ AMD HIP.}
\label{tab:portability}
\small
\setlength{\tabcolsep}{5pt}
\begin{tabular}{@{}l rr r@{}}
\toprule
Kernel & CUDA & HIP & Ratio \\
\midrule
Add (micro-op)  & 34,717  & 1,589  & 21.8$\times$ \\
Mul (micro-op)  & 34,758  & 1,586  & 21.9$\times$ \\
AXPY            & 43,950  & 1,119  & 39.3$\times$ \\
DOT             & 28,113  &   952  & 29.5$\times$ \\
Reduction       & 26,226  & 1,086  & 24.1$\times$ \\
Triad           & 29,495  & 1,708  & 17.3$\times$ \\
SpMV            & 54,100  & 1,745  & 31.0$\times$ \\
\bottomrule
\multicolumn{4}{@{}p{0.92\columnwidth}}{\footnotesize
A100 PCIe vs.\ RX~780M iGPU. The two columns were measured in separate
campaigns; primary A100 medians appear in Tables~\ref{tab:gpu-micro}
and~\ref{tab:kernels}. The ratios stem from the hardware gap, not the
quality of the code.}
\end{tabular}
\end{table}

\section{Related Work}
\label{sec:related}

Prior work differs in platform scope, precision model, and language runtime.

\textbf{CPU-only libraries.}
Hida \textit{et al.}~\cite{hida2001qd} introduced QD's \texttt{dd\_real} and
\texttt{qd\_real}, our principal CPU baselines; \ffx{4} Vec4 is
$4.2\text{--}6.6\times$ faster than \texttt{qd\_real}
(Table~\ref{tab:cpu-micro}). Mukunoki \textit{et al.}~\cite{mukunoki2020dsfp}
studied double--single precision--performance tradeoffs.

\textbf{Vendor-specific GPU libraries.}
CAMPARY~\cite{joldes2017campary} supports variable limbs on CUDA. \libname{}
matches it at \ffx{2} and is $1.4\text{--}1.6\times$ and
$1.1\text{--}1.6\times$ faster at \ffx{3} and \ffx{4}, respectively
(Table~\ref{tab:gpu-micro}). GQD~\cite{lu2010gqd} brought QD to CUDA, while
Zhang and Aiken~\cite{zhang2025sc} provided computer-verified error bounds for
the floating-point accumulation networks that underpin the branch-free
algorithms used in~\cite{multifloatsjl}.

\textbf{Reproducible accumulation.}
ExBLAS~\cite{iakymchuk2015exblas} uses long accumulators and compensation for
bitwise-reproducible BLAS rather than a higher working precision; both
approaches retain information discarded by native arithmetic.

\textbf{Ozaki-scheme tensor-core GEMM.}
Ootomo \textit{et al.}~\cite{ootomo2024gemm} use Ozaki splitting to emulate double-precision
GEMM on integer tensor cores~\cite{mukunoki2020dsfp}, which can exceed native
FP64 peak for suitable matrices but has conditioning-dependent accuracy. We instead
apply one $O(k^2)$ arithmetic model to AXPY, FFT, SpMV, reductions, and GEMM;
an Ozaki dispatch remains a promising extension.

\textbf{Julia \mfjl{}.}
\mfjl{}~\cite{multifloatsjl} implements the same algorithm family through
Julia and LLVM. Our GPU results remain within $1.2\times$, with competitive CPU
performance; \libname{} instead integrates as a header-only C++17 component.

\section{Conclusion}
\label{sec:conclusion}

We have shown one C++ source for CPU/GPU multi-float arithmetic. A100 \ffx{4}
addition sustains 15,700~MOps/s ($1.5\times$ CAMPARY), and \ffx{4} Vec4 is
$4.2\text{--}6.6\times$ faster than QD's \texttt{qd\_real}. On tested inputs,
addition and multiplication show 0~ULP error and square root stays within
17.9~ULP. Level~1 kernels reach approximately 93\% efficiency on 16 cores;
Kokkos overhead stays below 1.7\% on A100/GH200; and unchanged source runs
through CUDA/HIP. Future work targets Ozaki GEMM, SYCL, adaptive precision
and performance benchmarking on a binary black hole problem.

\section*{Acknowledgment}
M.V. thanks Sigal Gottlieb, Gaurav Khanna, Aaron Jarmusch and Connor Kenyon for their insights, and
the Kokkos and MultiFloats.jl developers. UMass Dartmouth provided NVIDIA
A100/GH200 access through MGHPCC Unity. Code/data and the artifact are in the Github repository
\url{https://github.com/manasvishal/MultiFloats.cpp}.

\clearpage
\balance


\begin{thebibliography}{00}

\bibitem{vishal2024toward}
M.~Vishal, S.~E. Field, K.~Rink, S.~Gottlieb, and G.~Khanna,
``Toward exponentially-convergent simulations of extreme-mass-ratio
inspirals: A time-domain solver for the scalar Teukolsky equation with
singular source terms,'' \textit{Phys.\ Rev.\ D}, vol.~110, no.~10,
pp.~104009, 2024.

\bibitem{vishal2025superconvergent}
M.~Vishal, S.~E. Field, S.~Gottlieb, and J.~Ryan, ``Superconvergent
discontinuous Galerkin method for the scalar Teukolsky equation on
hyperboloidal domains: Efficient waveform and self-force computation,''
\textit{General Relativity and Gravitation}, vol.~57, no.~7, pp.~104, 2025.

\bibitem{vishal2026pnp}
M.~Vishal, \textit{The PNP problem: precision and perturbation in black
hole dynamics: a dissertation in Engineering and Applied Science},
Ph.D.\ dissertation, University of Massachusetts Dartmouth, 2026.


\bibitem{bailey2005high}
D.~H. Bailey, ``High-precision floating-point arithmetic in scientific
computation,'' \textit{Computing in Science \& Engineering}, vol.~7, no.~3,
pp.~54--61, 2005.

\bibitem{fousse2007mpfr}
L.~Fousse, G.~Hanrot, V.~Lef\`{e}vre, P.~P\'{e}lissier, and P.~Zimmermann,
``MPFR: A multiple-precision binary floating-point library with correct
rounding,'' \textit{ACM Trans.\ Math.\ Software}, vol.~33, no.~2, pp.~13:1--13:15, 2007.

\bibitem{hida2001qd}
Y.~Hida, X.~S. Li, and D.~H. Bailey, ``Algorithms for quad-double precision
floating point arithmetic,'' in \textit{Proc.\ 15th IEEE Symp.\ Computer
Arithmetic}, 2001, pp.~155--162.

\bibitem{dekker1971floating}
T.~J. Dekker, ``A floating-point technique for extending the available
precision,'' \textit{Numerische Mathematik}, vol.~18, no.~3, pp.~224--242,
1971.

\bibitem{joldes2017campary}
M.~Joldes, J.-M. Muller, V.~Popescu, and W.~Tucker, ``CAMPARY: Cuda
multiple precision arithmetic library and applications,'' in \textit{Proc.\
5th Int.\ Congress on Mathematical Software (ICMS)}, 2016, pp.~232--240.

\bibitem{lu2010gqd}
M.~Lu, B.~He, and Q.~Luo, ``Supporting extended precision on graphics
processors,'' in \textit{Proc.\ 6th Int.\ Workshop on Data Management on
New Hardware (DaMoN)}, 2010, pp.~19--26.

\bibitem{multifloatsjl}
D.~K. Zhang and A.~Aiken, ``High-Performance Branch-Free Algorithms for
Extended-Precision Floating-Point Arithmetic,'' in \textit{Proc.\ Int.\ Conf.\
for High Performance Computing, Networking, Storage and Analysis (SC)}, 2025,
pp.~695--710, doi:~10.1145/3712285.3759876.

\bibitem{zhang2025sc}
D.~K. Zhang and A.~Aiken, ``Automatic Verification of Floating-Point
Accumulation Networks,'' in \textit{Proc.\ 37th Int.\ Conf.\ Computer Aided
Verification (CAV)}, 2025, pp.~215--237, doi:~10.1007/978-3-031-98682-6\_12.

\bibitem{trott2022kokkos}
C.~R. Trott \textit{et al.}, ``Kokkos 3: Programming model extensions for
the exascale era,'' \textit{IEEE Trans.\ Parallel Distrib.\ Syst.},
vol.~33, no.~4, pp.~805--817, 2022.

\bibitem{danalis2010shoc}
A.~Danalis \textit{et al.}, ``The scalable heterogeneous computing (SHOC)
benchmark suite,'' in \textit{Proc.\ 3rd Workshop on General-Purpose
Computation on Graphics Processing Units (GPGPU)}, 2010, pp.~63--74.

\bibitem{mukunoki2020dsfp}
D.~Mukunoki, T.~Ogita, and K.~Ozaki, ``Accurate and efficient computations
with Ozaki scheme in double-single-double floating-point format,''
\textit{Numerical Algorithms}, vol.~83, pp.~1431--1453, 2020.

\bibitem{iakymchuk2015exblas}
R.~Iakymchuk, S.~Collange, D.~Defour, and S.~Graillat, ``ExBLAS:
Reproducible and fast BLAS,'' in \textit{Proc.\ IEEE 22nd Symp.\ Computer
Arithmetic}, 2015, pp.~30--37.

\bibitem{rump1988algorithms}
S.~M. Rump, ``Algorithms for verified inclusions---theory and practice,''
in \textit{Reliability in Computing}, R.~E. Moore, Ed.\
Academic Press, 1988, pp.~109--126.

\bibitem{joldes2017bounds}
M.~Joldes, J.-M. Muller, and V.~Popescu, ``Tight and rigorous error bounds
for basic building blocks of double-word arithmetic,'' \textit{ACM Trans.\
Math.\ Software}, vol.~44, no.~2, pp.~15:1--15:27, 2017.

\bibitem{ootomo2024gemm}
H.~Ootomo, K.~Ozaki, and R.~Yokota, ``DGEMM on integer matrix multiplication
unit,'' \textit{Int.\ J.\ High Performance Computing Applications},
vol.~38, no.~4, pp.~289--306, 2024.




\end{thebibliography}
\end{document}